\documentclass{aastex}

\usepackage{emulateapj5}
\usepackage{epsfig}

\newcommand{\simle}{\mbox{$\stackrel{<}{_{\sim}}$}}
\newcommand{\simge}{\mbox{$\stackrel{>}{_{\sim}}$}}
\shorttitle{Interferometry with Keck~I}
\shortauthors{Tuthill et al.}

\begin{document}

\title{Michelson Interferometry with the Keck~I Telescope}

\author{P. G. Tuthill\altaffilmark{1,2}, J. D. Monnier\altaffilmark{2,3}, 
       W. C. Danchi\altaffilmark{3}, E. H. Wishnow\altaffilmark{3,4} and
       C. A. Haniff\altaffilmark{5}}

\altaffiltext{1}{Current Address: Chatterton Astronomy Dept, 
School of Physics, University of Sydney, NSW 2006, Australia}
\altaffiltext{2}{Current Address: Smithsonian Astrophysical Observatory 
MS\#42, 60 Garden Street, Cambridge, MA, 02138, U.S.A.}
\altaffiltext{3}{Space Sciences Laboratory, University of California at 
Berkeley, Berkeley,  CA, 94720-7450, U.S.A.}
\altaffiltext{4}{V Division, Lawrence Livermore National Laboratory,
7000 East Av., Livermore, CA, 94550, U.S.A.}
\altaffiltext{5}{ Astrophysics Group, Cavendish Laboratory,
Madingley Road, Cambridge, CB3 0HE, U.K.}

\email{gekko@physics.usyd.edu.au, jmonnier@cfa.harvard.edu, 
   wcd@ssl.berkeley.edu, wishnow@ssl.berkeley.edu, cah@mrao.cam.ac.uk}

\begin{abstract}
We report the first use of Michelson interferometry on the Keck~I
telescope for diffraction-limited imaging in the near infrared JHK and
L bands. 
By using an aperture mask located close to the f/25 secondary, 
the 10\,m Keck primary mirror was transformed into a separate-element, 
multiple aperture interferometer. 
This has allowed diffraction-limited imaging of a large number of 
bright astrophysical targets, including the geometrically complex 
dust envelopes around a number of evolved stars.  
The successful restoration of these images, with dynamic ranges in 
excess of 200:1, highlights the significant capabilities of sparse 
aperture imaging as compared with more conventional filled-pupil 
speckle imaging for the class of bright targets considered here.
In particular the enhancement of the signal-to-noise ratio of the 
Fourier data, precipitated by the reduction in atmospheric noise,
allows high fidelity imaging of complex sources with small 
numbers of short-exposure images relative to speckle.  
Multi-epoch measurements confirm the reliability of this imaging 
technique and our whole dataset provides a powerful demonstration 
of the capabilities of aperture masking methods when utilized 
with the current generation of large-aperture telescopes. 
The relationship between these new results and recent advances 
in interferometry and adaptive optics is briefly discussed.
\end{abstract}


\section{Introduction} 
Developments in detector technology and opto-electronic hardware over
the past decade have meant that real-time adaptive optical systems 
have now become a common feature of large ground-based optical and 
near-infrared telescopes (recent reviews may be found in 
Bonaccini \& Tyson 1998, Hardy 1998 \& Roddier 1999)
However, while adaptive optics has enjoyed considerable recent success,
other techniques that utilize post-detection data processing, rather 
than real-time compensation, have remained valuable for imaging at the 
very highest angular resolutions. 
The best known, and most straightforward of these to implement, is 
speckle imaging \cite{Lab70,weig91,NR96} in which sequences of 
short-exposures of a target and an unresolved calibrator are used to 
recover high-resolution maps beyond the natural seeing limit.  
Although this method in principle allows the recovery of images of 
arbitrary complexity, the difficulty of attaining an adequate 
signal-to-noise ratio has meant that it has mainly been confined to 
studies of binary stars (see, for example, Patience et al.~1998) 
and other astronomical sources with similarly simple geometries 
(though Weigelt et al.~1998 is a recent counterexample).

One solution to this signal-to-noise problem is to modify the pupil
geometry of the telescope using a mask so as to mimic the operation 
of a separated-element interferometer array such as the VLBA. 
This process can be considered as finding an optimal balance between 
the level of atmospheric perturbations, the number of photons, and 
the amount of structural information measured about the source --
all of which increase as the pupil area rises.  
When an aperture mask is being used, the data collection and analysis 
methods are similar to those utilized for speckle imaging, but with a 
reduction in the number of independent spatial frequencies measured, 
which is balanced by an improved signal-to-noise ratio on the 
data which are obtained.
This post-processing approach has been widely exploited at optical 
wavelengths where it has established itself as the only method by 
which reliable images of the surfaces of nearby stars at the 
diffraction limit have been recovered for ground-based telescopes 
(see, for example, Buscher et al.~1990; Wilson Dhillon \& Haniff~1997; 
Tuthill Haniff \& Baldwin~1999a).

In this paper, we report the first aperture masking experiments to
exploit the new generation of 10\,m-class telescopes. 
We have used the the Keck~I telescope with a variety of sparse 
multi-aperture pupil masks both to verify the signal-to-noise and 
calibration advantages of these pupil geometries, and to demonstrate 
the ability of this method to provide diffraction-limited imaging of 
resolved targets in the near infrared with excellent dynamic range. 
We have used multi-epoch measurements to establish the reliability of 
our imaging, and present near infrared maps of the highly-structured 
dust shells of a number of evolved stars at resolutions
exceeding 50\,milli-arcseconds.

\section{Experimental Design}

Aperture masking and conventional speckle interferometry share much in
common, including the ultimate goal of recovering the complex visibility
function of the target of interest (i.e.~the Fourier transform of its 
brightness distribution) at all spatial frequencies up to the telescope
diffraction limit. 
Generally, although Fourier amplitudes can be measured \cite{Fiz,Mich}, 
phases are scrambled by the atmosphere necessitating the 
use of an observable known as the closure phase \cite{Jenn58}.
More recently, closure phase concepts have been generalized into
the mathematical formalism of bispectral or triple correlation analysis
\cite{Loh83}.

Apart from the mask itself, the instrumental set-up required for masking
is almost identical to that for speckle, with sequences of short-exposure 
data frames being recorded at high magnification, allowing for the image
degradation caused by atmospheric turbulence to be removed during 
post-processing.
The principal difference between the two types of experiments is that in 
aperture masking the pupil geometry can be adjusted to optimize the 
signal-to-noise ratio (SNR). 
The implementation of a masking interferometer at the Keck~I telescope 
is discussed in more detail in the following sections.

\subsection{Optical Setup} \label{optics}

Unlike most earlier aperture masking experiments (e.g. Haniff et al. 1987;
Readhead et al. 1988),
which used screens placed in a small re-imaged pupil, masking of the Keck 
primary was achieved by placing large ($30$\,cm on a side hexagonal) aluminium 
masks directly in front of the f/25 infrared secondary mirror as shown 
in Figure~1. 
At this location, beams propagating to the detector are sufficiently 
well separated that masks can be treated as selecting discrete portions 
of the pupil, despite this not being a true pupil plane. 
With this design, however, the masks intercept radiation from the 
source twice: once on the way from the primary to the secondary, 
and a second time when traveling back towards the detector. 
As a consequence, masks selecting $N$ sub-apertures on the primary 
mirror in principle require $2N$ holes.
In order to accommodate this double-pass correctly, ray tracing software
was used to confirm that rays passing through each sub-aperture could 
indeed be traced back to discrete regions on the primary mirror.
The masks were fabricated of 3/16" aluminium sheet, and were mounted 
on a custom-built cylindrical post that protruded through the central 
hole of the f/25 secondary. 
At this location, the masks could be accessed during the night, 
permitting changes from one mask to another to optimize the configuration 
for a given source (this swap procedure took approximately 10\,min).
A typical desired pupil shape and the mask required to implement it 
are shown in Figure~\ref{mech_mask}.

After passing through the mask, the beams were focused on the Keck
facility Near-IR Camera, NIRC, using external magnifying optics 
(the so-called image converter; Matthews et al.~1996). 
The plate scale was 20.57 milli-arcsec/pixel on the 256$\times$256 
pixel InSb array; sufficient to Nyquist sample data collected in the 
K-band or longer wavelength bands. 
The observing wavelengths were selected from NIRC's
standard complement of interference filters which offered a range of
bandwidths (1~$\sim$~20\%) 
\medskip
\vbox{
\begin{center}
\epsfig{file=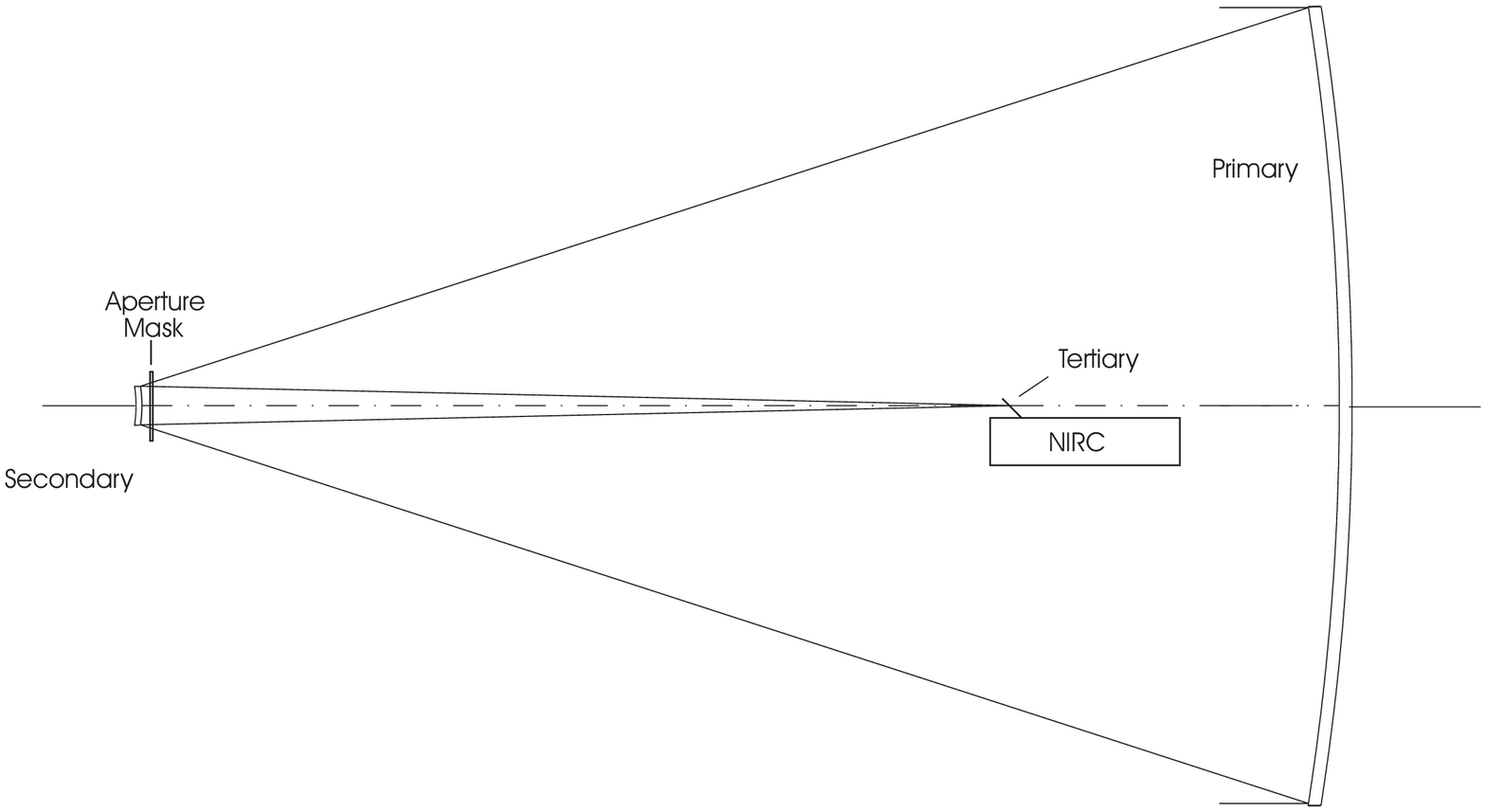,width=9.0cm}
\end{center}
\smallskip
{ \noindent \footnotesize \\ Fig 1.--- 
Optical ray-trace of starlight from the primary mirror of the
Keck telescope (right) to the Near InfraRed Camera (NIRC).
Aperture masks placed in front of the secondary mirror as shown must
be designed to take account of complications arising from the double 
light path, and the effects of the converging beam at this point in the
optical train.
 } }
\medskip
covering the 1~--~3.5\,$\mu$m region.

\setcounter{figure}{1}

\begin{figure*}[ht]
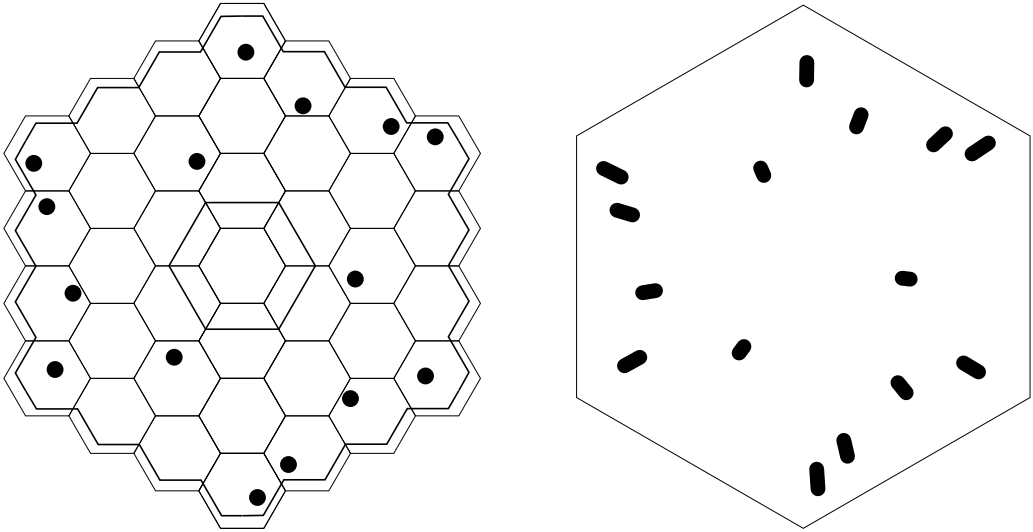

\begin{center}
\mbox{
\epsfig{file=golay15.eps,height=7cm}
\hphantom{XXXX}
\epsfig{file=golay15_mech.eps,height=7cm}
}
\end{center}
\caption{
A desired sparse pupil (left) and the aperture mask used to generate
it (right).  
The required pupil is shown as the set of black spots and is superimposed 
on a scaled version of the segmented 10\,m Keck primary mirror 
(hexagonal segments).  
The boundary of the undersized f/25 secondary mirror as projected on the
telescope primary is represented as the bold black line.  
The right-hand panel shows the corresponding aperture mask, ray-traced for
a location 60\,mm from the pole of the secondary mirror. 
In this case the elongated apertures allow the light to traverse both to 
and from the secondary on its way to the detector. 
The hexagonal mask plate is 300\,mm on a side. The central mounting hole 
for the mask is not shown in this figure.}
\label{mech_mask}
\end{figure*}

Although the masks were not cooled, this had little impact on the
experiment. 
At wavelengths shorter than the K band, the fast exposures 
(typically 140\,ms or less) ensured that there was little contribution 
from thermal emission. 
At longer wavelengths, thermal emission from the masks should have 
been a significant factor. 
However, the masks produced only a negligible increase in the thermal 
background, primarily because of the design of the NIRC magnifying optics. 
Because NIRC contains only a single cooled Lyot stop, when the image 
converter (Matthews et al.~1996) is in use, the cold stop is significantly
oversized. 
The array is exposed to room-temperature radiation (mostly from baffles
and other camera structures), so the presence of the masks had little 
effect on the thermal background, which was already dominated by ambient 
flux.

\subsection{Mask Design}
Most pupil mask designs used non-redundant configurations of sub-apertures, 
each with an effective size when projected onto the primary mirror of 
20~--~35\,cm.
This was tailored to be of order the seeing scale size, $r_o$. 
The lack of redundancy ensured that any Fourier component measured could 
be uniquely identified with a particular pair of sub-apertures, while
the small sizes minimized the effects of wavefront perturbations across 
each sub-aperture ($<1$ radian r.m.s). 
However, the segmented nature of the Keck primary mirror, and the 
undersized infrared secondary significantly complicated the task of 
locating the sub-apertures.
In particular when designing the masks, geometries where sub-apertures 
were crossed by the telescope spider or panel boundaries in the 
segmented primary mirror had to be avoided where possible. 
Designs were also driven by the desire that snapshot Fourier plane 
coverage be uniform and isotropic. 
The specific nature of these constraints meant that it was
not possible to exploit the results of Keto (1997) or Cornwell (1988),
both of whom explored optimum snapshot array configurations for radio 
interferometers.

\begin{figure*}[p]
\begin{center}
\mbox{
\epsfig{file=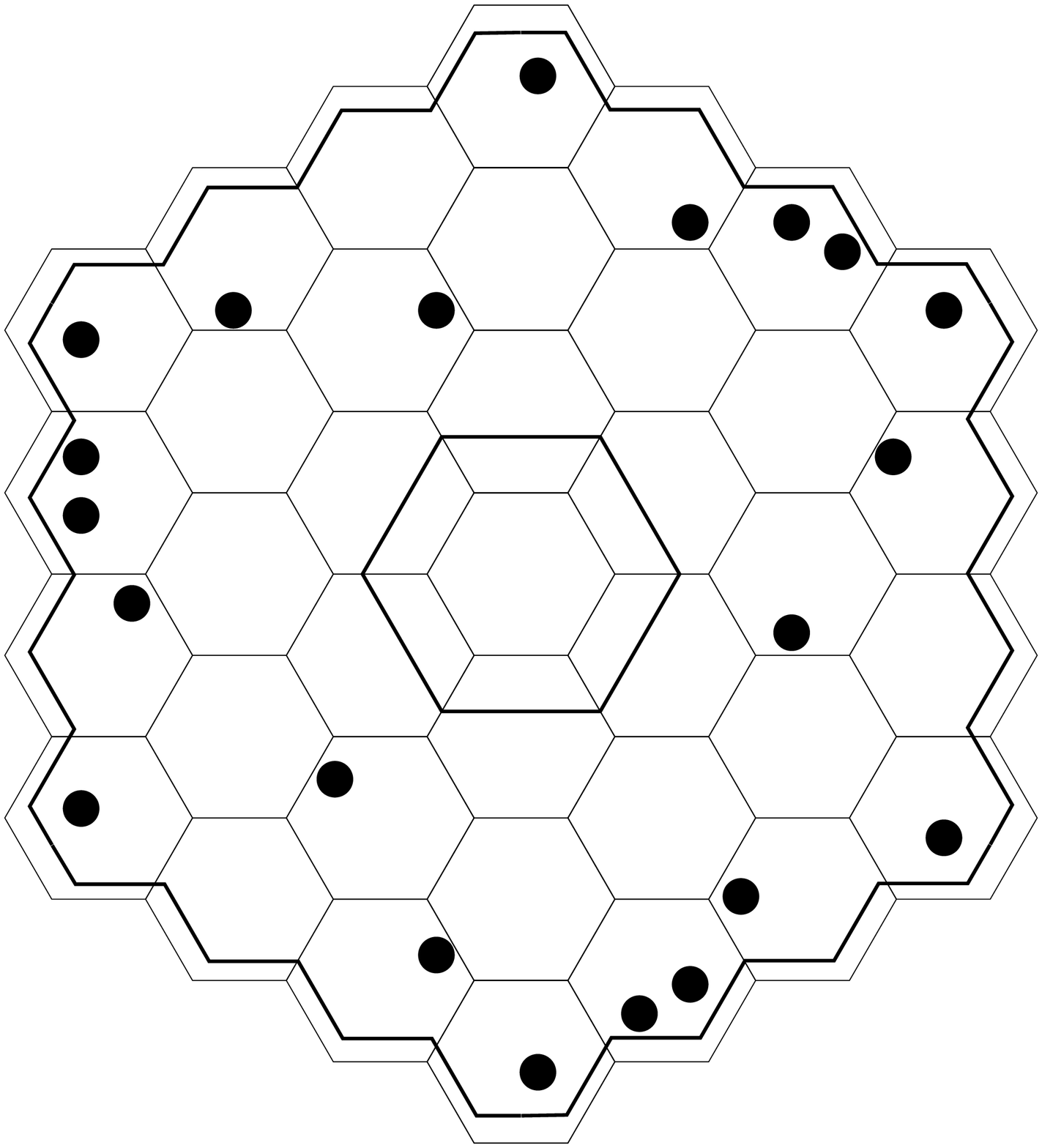,height=7.2cm}
\hphantom{XXXXXXX}
\epsfig{file=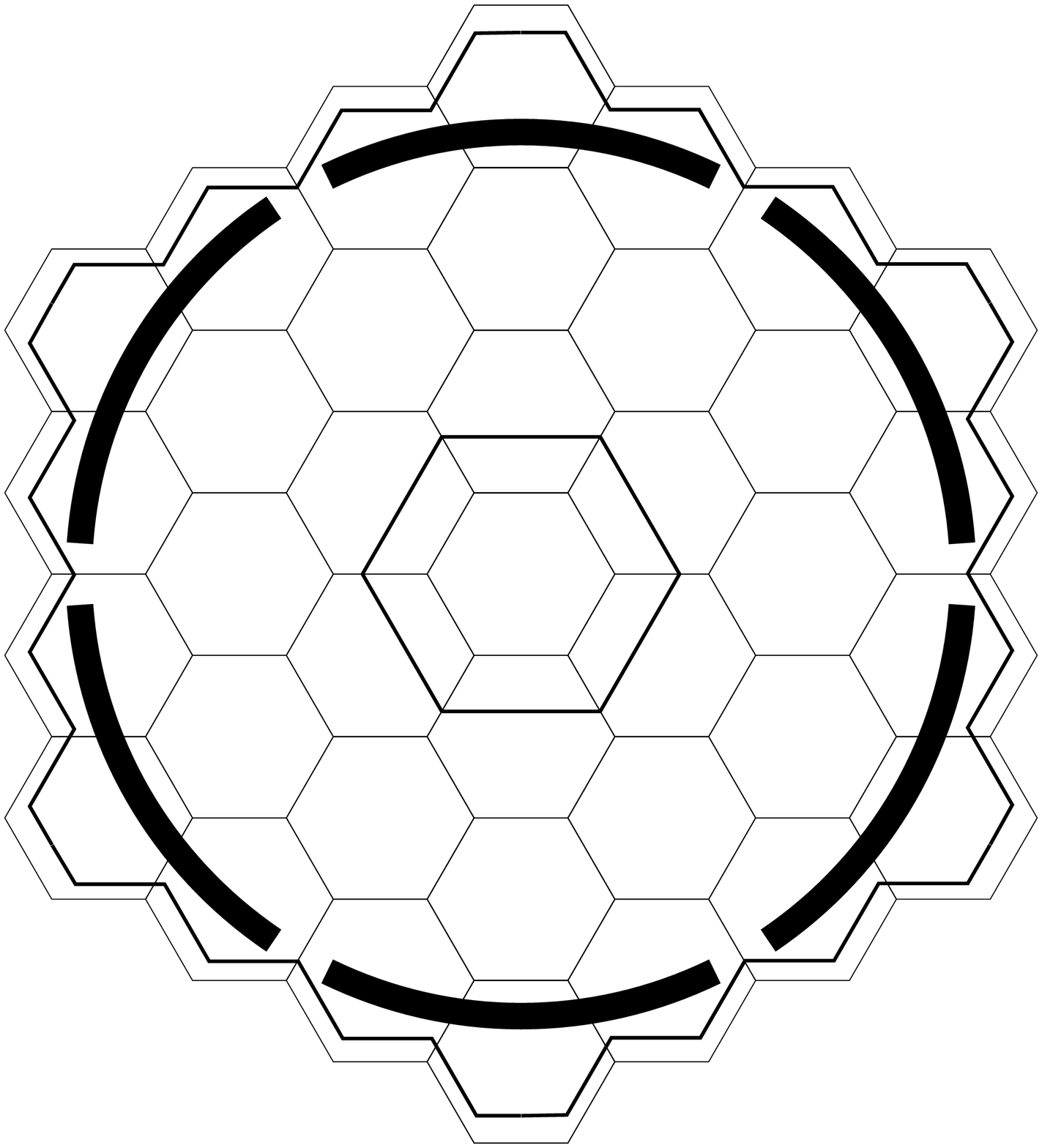,height=7.2cm}
 }
\mbox{
\epsfig{file=g21_spkle_neg.eps,height=7.2cm,angle=90}
\hphantom{XXXX}
\epsfig{file=ann_spkle_neg.eps,height=7.2cm,angle=90}
 }
\mbox{
\epsfig{file=g21_ps_neg.eps,height=7.2cm,angle=90}
\hphantom{XXXX}
\epsfig{file=ann_ps_neg.eps,height=7.2cm,angle=90}
}
\end{center}
\caption{
Aperture configurations (top), single exposure interferograms
(middle) and accumulated power spectra of 100 exposures (bottom)
for two different mask geometries.
Each mask is shown superimposed on the Keck primary.
Dark regions in the accumulated power spectra reveal the spatial
frequencies sampled by each pupil.
In the case of the 21-hole Golay mask (left panels) the Fourier
coverage is confined to a set of 210 independent frequencies,
whereas the annular mask (right panels) gives quasi-complete
coverage within a circular boundary.}
\label{mask_ps}
\end{figure*}

Mask designs were based on the approach of Golay (1970), in which 
3-fold symmetric spaces were searched for non-redundant array solutions
with compact and dense uv-coverage. 
Examples of arrays developed for 15- and 21-hole masks are shown 
in Figures~\ref{mech_mask} and \ref{mask_ps}. 
As can be seen in the 21-hole mask, the Fourier plane coverage has 
a densely filled core out to baselines of approximately 3\,m, and 
then adequate, but not isotropic coverage out to the edge of the
telescope pupil. 
Solutions with numbers of apertures as large as 36 were found, but for 
many experiments a 21-hole mask was more than adequate: this allowed 
the simultaneous measurement of 210 baselines and 1330 closure phases, 
yielding an excellent snapshot imaging capability.

\begin{figure*}[t]
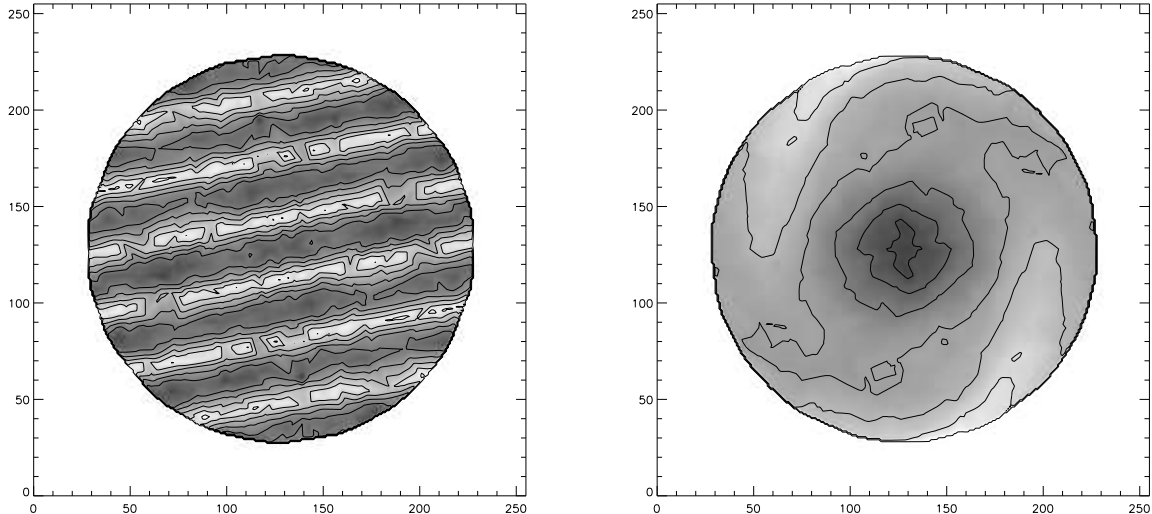

\begin{center}
\mbox{
\epsfig{file=binary_vis.eps,height=7cm,angle=90}
\hphantom{XXXX}
\epsfig{file=wr104_vis.eps,height=7cm,angle=90}
 }
\end{center}
\caption{
Examples of calibrated visibilities recovered from annular mask data.
Axes show the spatial frequency (UV) plane, with zero spatial frequency
translated to the center, while the boundaries of the figures are at
the nyquist sampling limit.
Note that the measurements have been interpolated to fill the Fourier
plane and so do not show the gaps in the Fourier plane coverage not
sampled by the mask.
The data in the left-hand panel are for the binary star $\alpha$~Com
and clearly display the expected sinusoidal modulations in visibility.
The right-hand panel shows the visibility function of the dusty
Wolf-Rayet star WR~104, a more resolved target with a large
asymmetric near-infrared envelope.
}
\label{two_d_vis}
\end{figure*}

Apart from defining the spatial frequencies measured by the telescope,
the aperture mask also served to limit the total amount of flux
collected.  
For observations of bright sources (M supergiants and Miras can 
have K magnitudes as high as $-4$\,mag) this was an important feature, 
as the use of the unobscured pupil would have saturated NIRC
in a small fraction of the minimum available exposure time, despite
the use of the narrowest available filters. 

For dimmer sources, non-redundant masks such as those described above 
did not transmit enough flux to overcome the array readout noise. 
The use of partially-redundant annular masks, as suggested by 
Haniff \& Buscher (1992), provides continuous Fourier plane coverage 
and enhanced throughput at the expense of only twofold redundancy.
This approach has been adopted here. 
An example of such a mask with a throughput of approximately 10\% 
is shown in Figure~\ref{mask_ps}, together with a short exposure 
interferogram and its power spectrum.
For sources with K magnitudes fainter than about 4\,mag, the SNR 
from masking interferometry was dominated by readout noise,
undermining the advantages offered by sparse pupils and 
necessitating the use of the full pupil.

One unusual feature of our non-redundant Golay masks is that they
allowed measurements of interference fringes that were sub-Nyquist
sampled. 
As mentioned in section~\ref{optics}, the pixel scale of NIRC
only allowed for Nyquist sampling of the longest available baselines
at wavelengths greater than 2\,$\mu$m. 
At shorter wavelengths, power at the corresponding spatial frequencies 
is ``aliased'' back into the power spectrum at lower spatial 
frequencies, and in most experiments, this signal will overlap and 
become confused with other shorter baseline signals. 
With the sparse pupils used here, however, these aliased signals were 
often mapped back onto otherwise unsampled frequencies and so 
measurements of the long baseline interference signals could still 
be made. 
In practice, the undersampling attenuated the signals significantly, 
and the recovery of useful data was only possible in cases where the 
signal-to-noise was initially very high.
Nevertheless, we have used this device to recover the J band
visibility functions of a handful of the brightest supergiants and
Miras at the highest spatial frequencies where the fringes were
sub-Nyquist sampled.

\section{Observing Procedure}
In view of the close relationship between aperture masking and
conventional filled aperture speckle interferometry, observations were
secured using a schedule very similar to that used in other high
resolution imaging experiments (see, e.g. Matthews et al.~1996). 
For each target 100 short exposure interferograms were collected, after
which a smaller number (usually 10) of identical exposures were
secured on nearby blank sky.  
The exposure times were limited by the array readout, with times 
of 140\,ms possible for the full $256 \times 256$ NIRC array.
This is longer than the typical atmospheric coherence time ($t_0$) 
expected at $2.2$\,$\mu$m ($\sim 40$\,ms) resulting in added 
atmospheric noise for the bispectral measurements. 
However, integration times of a few $t_0$ preserve significant fringe
power and actually deliver higher SNR in the photon-noise limited 
(faint source) regime \cite{Bush88}, while the closure phase is 
even more robust and can be measured with integrations of many 
$t_0$ \cite{Rhd88}.
Shorter exposure times could be obtained by reading out smaller 
sub-frames of the array, but in general the large sizes of the 
interferograms, resulting from the small dimensions of the pupil
sub-apertures, meant that sub-framing was problematic for reasons 
of data calibration and loss of field of view.

Calibration of the mean telescope-atmosphere transfer function was
performed by interleaving the 100\,source~+~10\,sky datasets with 
identical exposures of nearby similarly bright calibrator stars. 
Sets of these ``matched pairs" of data were secured for each source and
wavelength of interest, giving a total elapsed time for each 
complete observation of order ten minutes. 
Currently this time remains limited by the $\sim$10\% duty cycle of the 
real-time archiving software available at NIRC, which was not designed 
with this mode of operation in mind: custom designed hardware would 
likely increase the data collection rate by an order of magnitude.  
Despite the low duty cycle of the camera, the use of non-redundant 
pupils allowed high-signal-to-noise image reconstructions of resolved 
targets to be achieved with relatively small numbers of specklegrams 
(i.e.~as few as a hundred). 
This should be contrasted with the thousands typically used in 
filled-aperture speckle experiments where the atmospheric redundancy
noise limits the signal-to-noise per frame of the Fourier measurements
to unity even at high light levels.

Where possible, filter bandpasses of $\simle 5$\% were used in order 
to minimize the effects of the expected differences in spectral type
between the targets and their calibrator stars.
Without such precautions, the differing spectral shapes over the 
bandpass could affect the calibrated data.
The most interesting sources -- those with suspected well resolved
structure -- were observed at two well separated times during the 
night. 
This allowed independent checks to be made on the reliability of 
source structure determinations since spurious signals associated 
with the mask and detector array are not expected to rotate with 
the sky. 
Additional observations were also secured each night of a number of 
binary stars with well determined orbits to allow independent 
calibration of the detector orientation and scale and to assess the 
reliability of the subsequent data reduction pipeline.

\section{Data Reduction}

Analysis of the data followed standard methods for aperture masking
experiments and involved the accumulation of the power spectra and
bispectra of each set of interferograms. 
As a first step, each short exposure image was dark-subtracted, flat 
fielded, and cleaned of pattern noise arising from transients in the 
readout electronics and variations in the behavior of the four 
different readout amplifiers.
Images were subsequently windowed with a two-dimensional Hanning 
function tapered to zero so as to eliminate edge effects.
Power spectra could now be computed frame by frame from the squared
modulus of the Fourier transform.
Stellar fringe signals appeared as power at discrete locations in such 
spectra, with the origin occupied by a peak whose height was proportional
to the squared flux in the frame, while the remaining areas were filled
with a signal caused by a combination of photon and readout noise
(for illustration, see Figure~\ref{mask_ps}).

The noise power level, which would otherwise bias the measurements, could 
be obtained by averaging over those regions where no stellar signal was 
expected (usually, but not always, at the edges of the power spectra).
Having subtracted off the noise bias, squared visibilities (Fourier 
amplitudes) for the stellar interference signal were found by taking the 
ratio of the power at the spatial frequency of the fringes to the power at 
the origin, and then normalizing with the corresponding signal from the 
calibrator spectrum. 
Error estimates were derived from the spread in values amongst each 
ensemble of 100 exposures. 
Some typical results from this procedure are shown in 
Figure~\ref{two_d_vis}.

Fourier triple products, or bispectral data, could also be computed
for each frame.
Closure phase information was recovered from the argument of the
complex bispectral data accumulated over the ensembles of short 
exposures.
Again the measured closure phases for the reference stars 
were used to calibrate the measurements of the targets, however
such corrections, indicative of optical aberrations in the 
telescope/camera, were small (\simle few degrees).
Unlike filled aperture speckle experiments, the use of an aperture mask 
ensured that only a very small subset of the full bispectral hypervolume 
needed to be accumulated dramatically reducing computational and 
data storage requirements. 
A standard workstation was adequate for all data processing.

Having obtained sets of calibrated Fourier amplitudes and closure
phases, diffraction-limited images were recovered using standard radio
astronomical self-calibration methods. 
These techniques, originally developed for dilute phase-unstable 
radio arrays such as the VLBA, are readily transferable to this 
application since our data, i.e.~sampled Fourier amplitudes and closure 
phases, are almost identical to those delivered by arrays such as the VLBA. 
The imaging results reported here were obtained using a ``Maximum Entropy 
Method'' based implementation of self-calibration (Gull \& Skilling 1984, 
Sivia 1987), but in all cases reconstructions from CLEAN-based
methods (H\"{o}gbom 1974) gave similar results. 
In many instances the extraction of quantitative information was achieved
in a more robust and precise fashion from model-fitting directly to the 
Fourier data than from mapping, particularly when source structure was
relatively simple and/or partially resolved.

\begin{figure*}[ht]
\begin{center}
\epsfig{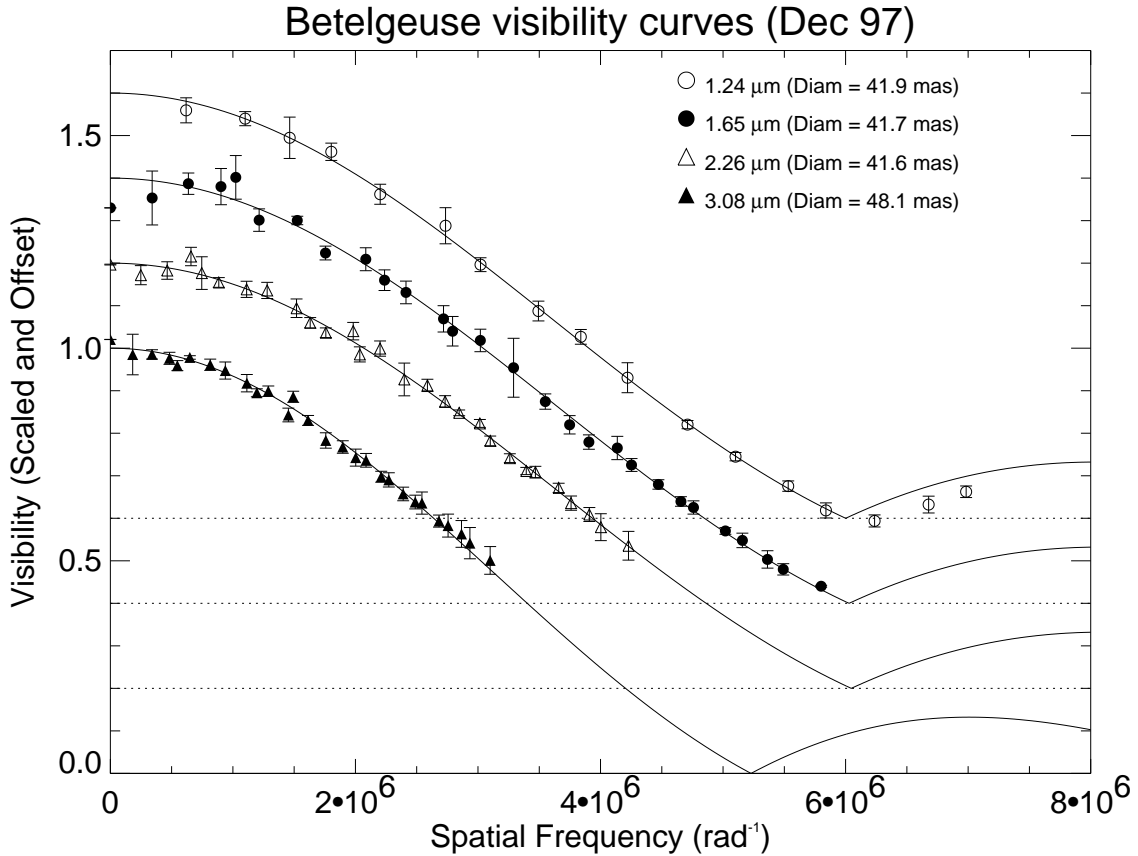}
\end{center}
\caption{Azimuthally averaged visibility amplitudes for $\alpha$~Ori
as measured in 1997 December at four different wavelengths.
In each case the best-fit uniform disk model (see key for diameters)
is shown overplotted as a solid line.
To avoid confusion, the data at the different wavelengths have been
offset by fixed amounts.}
\label{betelvis}
\end{figure*}

\section{Wavefront Coherence}

Since the success of interferometric imaging is critically
dependent on the coherence properties of the incoming wavefront,
experiments such as the ones reported here can in principle yield
valuable information on the stability and aberrations introduced by
mechanical deformation of the telescope, and by fluctuations in the
atmosphere.  
The former of these is particularly relevant in consideration of
the performance of the segmented primary mirror.
Quantitative investigations require a numerical simulation of
the phase irregularities introduced by both the atmosphere and 
telescope pupil. 
Although we have performed such computations, a full discussion 
of this work lies beyond the scope of the present report. 
Instead, a number of general conclusions and difficulties experienced,
some of which may be peculiar to the Keck, are outlined below.

The most obvious manifestation of optical aberrations related to the
segmented primary is the loss of spectral power (or fringe visibility), 
implying decorrelation of the wavefronts, at spatial frequencies which 
can be traced back to the locations of the edges of the hexagonal primary
panels. 
This effect, sometimes referred to as ``print-through'' from the segmented
pattern which appears in the power spectra, has also been seen by workers
undertaking full-pupil speckle observations (Ghez 1997). 
We have studied this effect in two ways.
First, numerical simulations involving turbulence degraded wavefronts 
have verified that significant phase discontinuities (\simge 0.5\,$\lambda$)
at the segment boundaries do indeed result in a loss of spectral power
which mimics the observations.
Second, experimental confirmation of this effect has been obtained
by deliberately displacing selected mirror segments in small increments 
with respect to their neighbors.
The overall finding of these studies is that between 1995 through 1997, 
the Keck~I primary was poorly phased when working in the infrared, with
most segment edges exhibiting \simge 0.5\,$\lambda$ phase steps.
This can probably be traced to sub-optimal performance of the ``malign''
alignment procedure.
In 1997, however, the primary mirror alignment procedures \cite{Troy98,Chanan98}
were refined and since then, phase distcontinuities, 
while still present, have been greatly ameliorated.

A further problem we have experienced that is possibly related to the
segmented primary concerns the anomalously low visibilities measured
at long baselines.
Measured point-source visibilities were lower by a factor of 2\,$\sim$\,3
than the values expected on the basis of numerical simulations.
One likely candidate for this loss is the active control system responsible 
for maintaining the relative alignment between segments. 
Significant oscillation or jitter of the system is known to occur 
(Wizinowich 1999), which perturbs the segments at frequencies of tens of 
Hertz -- rapidly enough to blur the fringes and introduce a loss in 
visibility comparable to that we have observed.

\begin{figure*}[ht]
\begin{center}
\epsfig{file=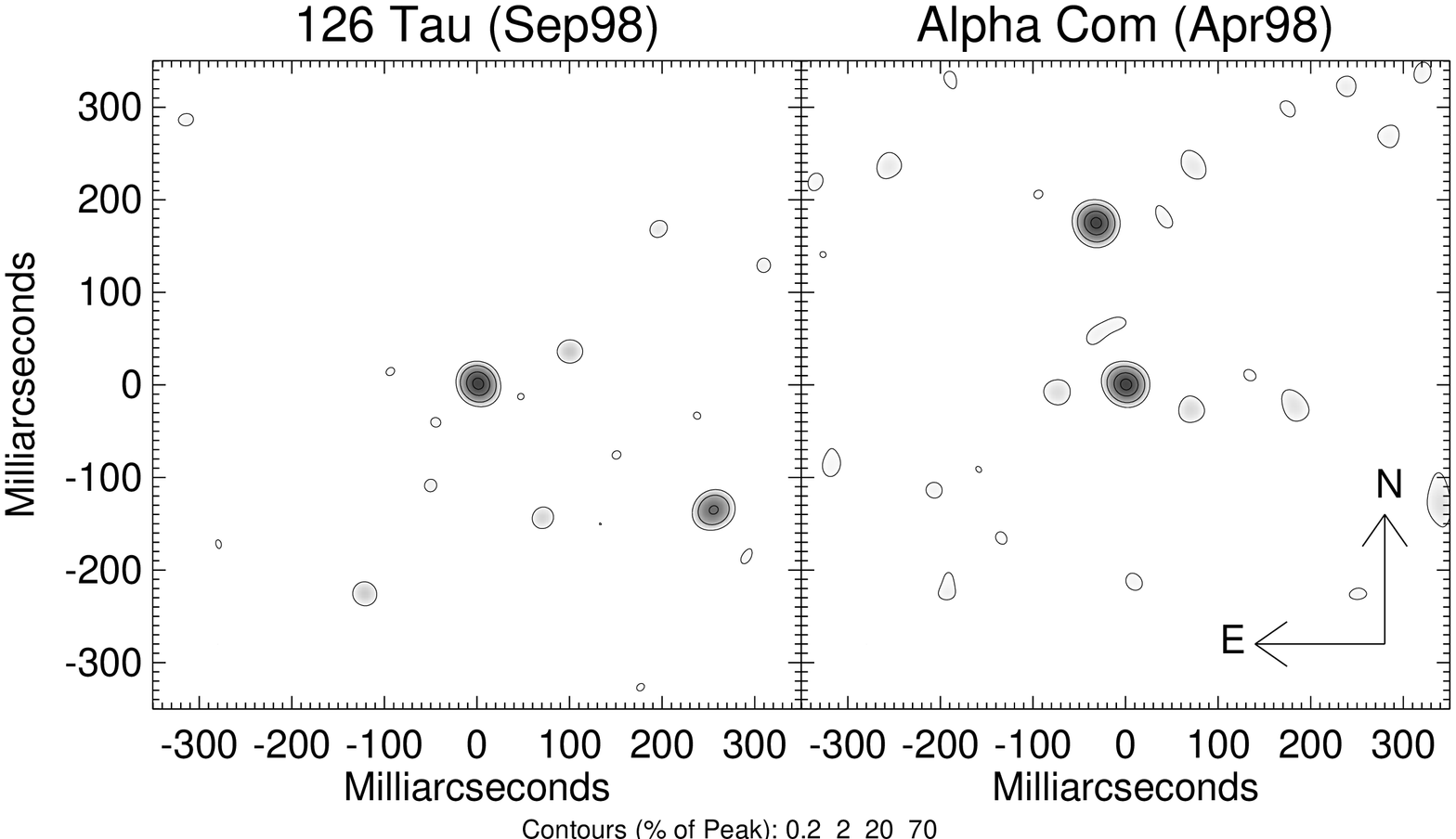,height=17.0cm,angle=90}
\end{center}
\caption{
Reconstructed K-band images of the binary stars 126~Tau and
$\alpha$~Com.
The logarithmic contour levels, at 0.2\%, 2\%, 20\% and 70\% of the
peak, were chosen to highlight the very small noise features in each map.
Although the binary separations for these stars are only 287 and
177\,mas respectively, they are easily separated at the diffraction
limit of the Keck telescope.
}
\label{binaries}
\end{figure*}

Perhaps most damaging of the limitations encountered was the
difficulty in ensuring the intermediate term ($\sim$\,10\,min)
stability of the optical transfer function necessary for reliable
calibration of the measurements.  
Mis-calibration effects, introduced where there had been a change in the 
seeing-averaged system transfer function between measurements of source
and calibrator, were the major obstacle in recovering high fidelity 
image reconstructions.
One can envisage many sources of this problem.
Without a doubt, changes in the local seeing could have occurred over 
the \simge\,4\,min time lag between source and calibrator observations. 
Although the use of an aperture mask is known to limit the sensitivity
of spatial interferometry to seeing variations, some mis-calibration 
must be traced to this. 
Alternatively, any flexture of the telescope structure or changes in the 
phasing of the primary mirror could contribute to a non-stationary 
transfer function. 
In practice, the reference stars used were invariably at different 
elevations in the sky (worst cases could be greater than ten degrees 
away), and inevitable changes in the telescope wind loading and 
temperature all will have contributed to the difficulty of maintaining 
tolerances for interferometric observations.
Further work to identify and remedy this calibration problem would 
dramatically enhance the precision of this high spatial resolution 
experiment.

\section{Results and Discussion}

Observing in the near-infrared with the 10\,m baselines available at the 
Keck yields an angular resolution appropriate for large targets.
The excellent snapshot Fourier coverage of the telescope provides a 
densely sampled a stellar visibility function with orders of magnitude 
greater efficiency than a separated element array with a small number
of stations. 
The ability to secure data rapidly and through a wide range of narrow 
bandpasses has been particularly useful for investigating the 
atmospheric structure of cool supergiants.  
Stellar diameters are common science targets in high-angular
resolution astrophysics (see, e.g.~van Belle et al.~1999 for some
recent results). 
Figure~\ref{betelvis} shows the azimuthally averaged visibility function 
of Betelgeuse ($\alpha$~Ori) at four different wavelengths as measured in 
1997 December. 
Although a uniform disk model appears to be quite a good fit to the 
data at all the measured wavelengths, the star appears to exhibit an 
anomalously large diameter at $3.08$\,$\mu$m ($\delta\lambda = 0.10$\,$\mu$m), 
probably related to the presence of atomic or molecular absorption within 
the bandpass changing the optical depth of the stellar atmosphere.

In cases where the sources were significantly resolved, very high
quality images could be obtained. 
Figure~\ref{binaries} shows maps of the binary stars 126~Tau and 
$\alpha$~Com recovered from 100 interferograms. 
For both maps the dynamic range achieved, as measured by the ratio of 
the noise in the map to the peak intensity, was considerably better 
than 0.5\% or 1:200.
The nonlinear contour levels in the plots of Figure~\ref{binaries}
were chosen to highlight features at the level of the noise.
The ultimate limitations to the dynamic range can be traced back to
systematic miscalibration of the visibility amplitudes and poor 
handling of noise on the closure phases by the mapping software.
Insufficient Fourier sampling and low signal flux -- problems which are
exacerbated by the use of a mask -- were rarely an important contribution 
to the mapping error budget.

\begin{figure*}[ht]
\begin{center}
\epsfig{file=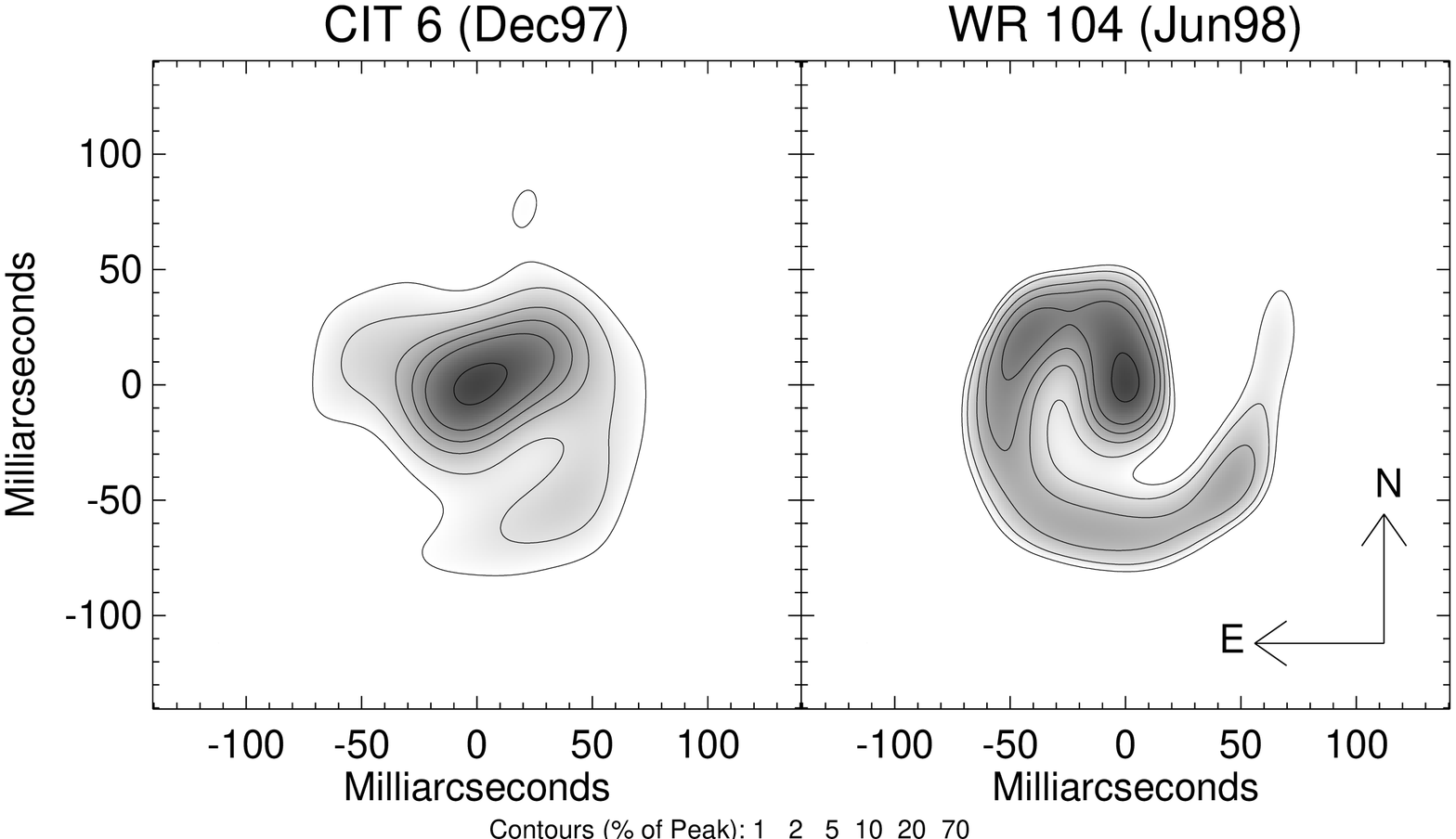,height=17.cm,angle=90}
\end{center}
\caption{
Diffraction-limited images of the evolved Carbon star CIT\,6 (left) and
the dusty Wolf-Rayet star WR\,104 (right) reconstructed from K-band
measurements taken with the 15-hole Golay and annular mask respectively
(see Figures~\ref{mech_mask} and \ref{mask_ps}).
For asymmetric objects such as these, the recovery of high-quality
phase information is crucial in recovering the true flux distribution.
}
\label{prettymaps}
\end{figure*}

More powerful arguments for the use of sparse pupil geometries in near
infrared speckle imaging are provided by our results for complex
resolved sources. 
Figure~\ref{prettymaps} shows two such images of the dust enshrouded 
Carbon star, CIT\,6, and the IR-bright Wolf-Rayet star WR\,104. 
Both of the sources are clearly resolved at the tens of millarcsecond 
scale, and their complex structures, which have been confirmed through 
independent repeated observations (see, e.g., Tuthill et al.\ 1999b), 
demonstrate the unique facility that the combination of interferometric 
methods with the large Keck primary provides. 
Both maps have dynamic ranges better than 100:1, and are of comparable 
quality to the images routinely produced by modern radio VLBI arrays.
Illustrations of the high level of repeatability for maps taken 
over separate epochs can be found in Monnier et al.~(1999a).

Scientific results from the masking program at Keck have encompassed
a range of astronomical topics from measurements of stellar 
photospheres  (e.g. Monnier et al.\ 1997; Tuthill Monnier \& Danchi 1998b;
Tuthill et al.\ 2000)
to imaging circumstellar dust shells in evolved stars
(e.g. Tuthill et al.\ 1998a,1999c; Danchi et al. 1998; 
Monnier et al.\ 1999a) and enshrouded Wolf-Rayets (Tuthill et al.\ 1999b,c;  
Monnier Tuthill \& Danchi 1999).
As is discussed further in Monnier (1999), the parameter space addressed 
by this experiment -- bright objects ($m_k \simle 4$\,mag) with resolvable
structure on 10\,m baselines -- is a particularly rich one since the 
combined resolution and magnitude limits are fortuitously matched for
targets containing hot astrophysical dust at around $\sim 1000$\,K.
Further discussion of the program stars and techniques may be found
in Tuthill et al.\ (1998c) and Monnier (1999).

In contrast to the situation only decades ago, astronomers are now armed
with a number of different techniques all aimed at overcoming the
seeing limit in astronomical observations. 
These include aperture masking, separate element interferometry, 
speckle interferometry, observations from space, and adaptive optics (AO), 
with the distinctions between these areas becoming increasingly blurred.
Non-redundant, or partially-redundant masking occupies a particular niche 
in this parameter space, and has been successful for bright objects 
resolvable with relatively modest baselines.
The primary competing technologies here are speckle interferometry
and adaptive optics, which we discuss in turn below.

Although there has been much debate on the relative merits of masking
versus speckle interferometry, there is agreement that for the faintest
objects (where ``faint'' is ultimately defined as a small number of
photons in a coherent patch per coherence time, but in practice is
usually governed by readout noise in the IR array detector before this 
limit is attained) then speckle interferometry is the superior choice.
For bright sources, there is little doubt that masking offers dramatic 
signal-to-noise advantages for discrete measurements of Fourier 
amplitudes and closure phases (e.g. Roddier 1987).
However, filled-pupil speckle interferometry is capable of recovering far
greater volumes of data, completely filling the bispectral volume, albeit
at low signal-to-noise.
A number of studies have addressed the comparison of non-redundant versus
filled pupil from theoretical, numerical simulation, and observational 
approaches \cite{Rhd88,HB92,BH93}, and have concluded
that there are clear regions where masking can outperform filled-pupil
techniques in recovering diffraction limited images of arbitrary celestial
objects.
In practice, superb images have been recovered using both techniques
(for examples of recent speckle images, see Weigelt et al.\ 1998).
A choice of which technique should be used is often also driven by more
mundane considerations, which in our case included saturation of the 
camera on bright stars and the availability of a well characterized and
mature image reconstruction code, both of which favor a masking strategy.

In making a comparison of the relative merits of adaptive optics
a separate set of considerations present themselves.
Many of the masking program stars, in addition to being bright, also 
exhibit a compact core and therefore might be thought ideal targets
for natural guide-star adaptive optical systems.
However, a strong note of caution needs to be sounded against the 
optimistic projection that such systems will render post-processing
techniques such as masking or speckle rapidly obsolete.
Obtaining true diffraction-limited images from current AO data requires 
careful deconvolution of the point-spread function (PSF). 
It has been our experience \cite{vycma} that the PSF of an AO system is 
difficult to fully characterize, and worse, is not stationary when the 
telescope is moved to a calibration star which will usually be of 
different brightness, elevation and spectrum.
In his recent review of AO, Ridgway (1999) notes that the problem of 
erratic PSF artifacts, which can easily masquerade as genuine source 
structure, is endemic to current-generation AO systems.
For all its unappealing appearance, an uncompensated speckle cloud 
has the advantage that it results from simpler underlying processes --
the atmosphere and telescope optics only -- with the result that
(in the high light limit) it is easier to calibrate.
We hasten to point out that AO offers numerous and dramatic advantages 
across a wide spectrum of observational problems (e.g. Ridgway 1999);
our discussion here is limited only to bright stars with structure
at the highest angular resolutions.

\section{Conclusions}

Results from the first aperture masking experiment performed on a 10\,m
class telescope are presented.
A suitable choice of non- or low-redundancy pupil geometries has been
found to dramatically improve the signal-to-noise on recovered bispectral
data.
Reliable images with complex and asymmetric structure at the diffraction 
limit have been routinely produced in the near-infrared JHK and L bands.
With dynamic ranges in excess of 200:1, and demonstrated repeatability
of map structure over multiple observing epochs, the expected advantages
of sparse-aperture interferometry for bright targets have been confirmed. 
In a comparison of aperture masking, full-pupil speckle, and adaptive 
optics, the most reasonable conclusion appears that each technique has
regions of the parameter space of source-brightness and spatial structure 
where it offers superior performance.
The existence of such complementary observational techniques will certainly
be beneficial in addressing a range of problems in high resolution astronomy, 
with masking being at its most effective for the brightest objects at the 
highest angular resolutions.
The robust reconstruction of complex brightness distributions from sparsely
sampled Fourier data augers well for the future of the next generation 
of separate-element ground-based imaging arrays with baselines in excess 
of an order of magnitude larger than those available here.

\acknowledgments

The Authors would like to thank Gary Chanan and Mitchell Troy for their
help with performing the segment phasing experiments. 
We would also like to thank Everett Lipman, Charles Townes, Peter
Gillingham, and Terry McDonald, all of whom have contributed to the
success of our endeavors.
Devinder Sivia kindly provided the maximum entropy mapping program ``VLBMEM'', 
which we have used to reconstruct our diffraction limited images. 
This work has been supported by grants from the National Science Foundation 
(AST-9321289 and AST-9731625).
CAH is grateful to the Royal Society for financial support.
EHW was supported in part under the auspices of the University Relations
Program at Lawrence Livermore National Laboratory using UCDRD funds,
and by the U.S. Department of Energy at LLNL under the contract no.
W-7405-ENG-48.


\clearpage

\end{document}